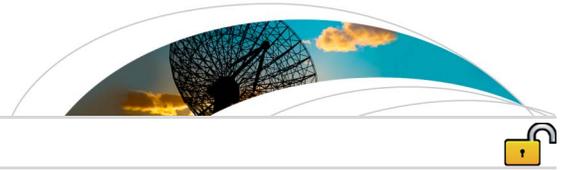

# Variable phase propagation velocity for long-range lightning location system

Zhongjian Liu[1], Kuang Liang Koh[1], Andrew Mezentsev[2], Sven-Erik Enno[3], Jacqueline Sugier[3], and Martin Füllekrug[1]

[1]Centre for Space, Atmospheric and Oceanic Science, University of Bath, Bath, UK, [2]Birkeland Centre for Space Science, University of Bergen, Bergen, Norway, [3]Remote Sensing (Lightning, Clouds, Aerosols), and Aircraft Based Observation R&D, Met Office, Exeter, UK

**Abstract** The electromagnetic wave propagation velocity at low radio frequencies is an important input parameter for lightning location systems that use time of arrival (TOA) method. This velocity is normally fixed at or near the speed of light. However, this study finds that the radio waves from two submarine communication transmitters at 20.9 kHz and 23.4 kHz exhibit phase propagation velocities that are ~0.51% slower and ~0.64% faster than the speed of light as a result of sky wave contributions and ground effects. Therefore, a novel technique with a variable phase propagation velocity is implemented for the first time in the TOA method and applied to electric field recordings with a long-baseline lightning location system that consists of four radio receivers in western Europe. The lightning locations inferred from variable velocities improve the accuracy of locations inferred from a fixed velocity by ~0.89–1.06 km when compared to the lightning locations reported by the UK MetOffice. The normal distributions of the observed phase propagation velocities in small geographic areas are not centered at the speed of light. Consequently, representative velocities can be calculated for many small geographic areas to produce a velocity map over central France where numerous lightning discharges occurred. This map reflects the impact of sky waves and ground effects on the calculation of lightning locations as a result of the network configuration. It is concluded that the use of variable phase propagation velocities mitigates the influence of sky waves and ground effects in long-range lightning location networks.

## 1. Introduction

Lightning is the strongest natural electromagnetic radiation source, and it emits electromagnetic energy in the frequency range from ~4 Hz to about ~300 MHz or more [*Rakov and Uman*, 2003; *Rison et al.*, 2016]. Both ground-based [e.g., *Rakov*, 2013] and space-based [e.g., *Christian et al.*, 2003; *Goodman et al.*, 2013] lightning location systems have been developed using a number of different techniques since the twentieth century. Ground-based lightning location systems can be implemented using different frequency ranges and lengths of baselines. The three most popular methods are magnetic direction finding (MDF) [e.g., *Horner*, 1954, 1957; *Krider et al.*, 1976], time of arrival (TOA) [e.g., *Lee*, 1986; *Füllekrug and Constable*, 2000; *Dowden et al.*, 2002], and interferometry [e.g., *Mardiana et al.*, 2002; *Mazur et al.*, 1997; *Stock et al.*, 2014; *Rison et al.*, 2016]. The application of these methods is driven by the development of novel technology. For example, the national lightning detection network (NLDN) upgrade to use TOA in combination with MDF, i.e., the improved accuracy from combined technology (IMPACT) [*Cummins et al.*, 1998] was enabled by the availability of GPS timing. Similarly, the interferometric method was recently developed for low-frequency radio waves [e.g., *Lyu et al.*, 2014; *Stock et al.*, 2015] as a result of the advancement of digital processing technology. The fact that many researchers contribute essential novel results in the field of lightning location indicates that the improvement of lightning location system accuracy and detection efficiency is an important and active research area [e.g., *Rodger et al.*, 2004; *Said et al.*, 2010; *Dienderfer et al.*, 2009; *Mallick et al.*, 2014; *Wang et al.*, 2016; *Sun et al.*, 2016].

All lightning location techniques are based on the recordings of radio waves that propagate from the lightning discharges to the receivers. Radio wave propagation depends on numerous factors such as the frequency, ionospheric height, terrain, and ground conductivity [e.g., *Schonland et al.*, 1940; *Barr et al.*, 2000]. This natural variability results in uncertainties of the computed lightning locations. An improved understanding of the underlying physics of lightning location uncertainties enables novel opportunities toward an improvement of the TOA method [*Cummins et al.*, 2010]. The correction of timing errors caused by propagation effects are important for each lightning detection sensor in a network [e.g., *Honma et al.*, 1998;

*Schulz et al.*, 2016]. There have been many studies of the time delay and amplitude attenuation due to electromagnetic wave propagation over different ground conductivities [e.g., *Cooray and Lundquist*, 1983; *Caligaris et al.*, 2008; *Cooray*, 2009] and terrain [*Li et al.*, 2015, 2016a, 2016b]. These effects result in a wave propagation velocity that has a direct influence on the determination of lightning locations and might enable an improvement of the geolocation accuracy of lightning discharges [e.g., *Jean et al.*, 1960; *Steele and Chilton*, 1964; *Chapman et al.*, 1966; *Dowden et al.*, 2002].

Long-range lightning location systems commonly use the group velocity, which is always less than or equal to the speed of light. The group velocity (Vg = $\delta\omega/\delta k$) is the velocity of a group of waves within an amplitude envelope. The phase velocity (Vp = $\omega/k$) determines the change of phase at a given location of one radio frequency component. For long-range lightning location systems (~500 km), the received radio waves are normally a mixture of the ground wave and sky waves, i.e., ionosphere hops. The contribution of the sky wave results in an elevation angle of the incident wave, which is seen by a receiver array as the slowness of the wave [e.g., *Füllekrug et al.*, 2015]. The wording "slowness" is used in seismology—it is suggested here to use the wording "phase propagation velocity" for radio waves. The phase propagation velocity is affected by the wave arrival elevation angle ($\theta$) and the wave arrival velocity (Vg) that is influenced by the ground and ionospheric parameters. This phase propagation velocity (Vg/cos$\theta$) is meant to be the velocity as it appears to a receiver array considering the elevation angle of the incident wave. For example, the phase propagation velocity will be infinite if the wave arrives with a ~90° elevation angle at the receiver array. The phase propagation velocity will be the group velocity if the wave arrives horizontally with a 0° elevation angle. This velocity with a given elevation angle is suitable for a 2-D lightning location calculation because the propagation distances are normally inferred from ground paths.

The scientific literature reports controversial results on the question whether the phase velocity in very low frequency (VLF) is larger or smaller than the speed of light in the Earth-ionosphere waveguide [e.g., *Jean et al.*, 1960 and *Steele and Chilton*, 1964]. To resolve this controversy, it would be helpful to investigate VLF transmissions with the aim to infer the phase propagation velocity that results from the effects of sky and ground waves in a steady state transmission. This contribution of sky waves and ground effects may also influence long-range lightning location system performance, such as the ATDnet of the UK MetOffice [*Bennett et al.*, 2011]. The experimental long-range lightning location system with >500 km receiver separation uses the frequency range between 5 kHz and 15 kHz, which contains a significant portion of the electromagnetic energy deposited by lightning return strokes [*Füllekrug et al.*, 2013b]. A novel technique using variable phase propagation velocity is proposed here to mitigate the impact of sky waves and ground effects on the determination of lightning locations and to study the spatial variability of the phase propagation velocity in the given frequency range over central France.

## 2. Long-Baseline Low-Frequency Radio Receiver Array

An experimental long-baseline lightning location system was deployed in western Europe and recorded electric field strengths continuously from 15:00 until 24:00 (UTC) on 8 August 2014, when a mesoscale convective system developed over central France.

The array consists of four radio receivers distributed over the UK and France (Figure 1, left). These receivers are located in Bath (BTH), Orleans (ORL), Lannemezan (LMZ), and Rustrel (RST). Each receiver continuously records electric field strengths between ~4 Hz and ~400 kHz with a sampling frequency of 1 MHz [*Füllekrug*, 2010]. This radio receiver was originally designed for sprite and lightning research [e.g., *Füllekrug et al.*, 2006, 2013a, 2014; *Mezentsev and Füllekrug*, 2013; *Soula et al.*, 2014]. The timing accuracy of the GPS clock is ~20 ns and enables the array to locate lightning sources and to study the recorded radio signals with high temporal resolution.

## 3. Phase Propagation Velocities of Radio Waves From VLF Communication Transmitters

Low-frequency radio waves are a mixture of ground waves and sky waves such that the waves arrive at a radio receiver with a certain elevation angle [e.g., *Füllekrug et al.*, 2015]. In order to assess the impact of the elevation angles of sky waves and ground effects, the phase propagation velocities of radio waves from well-known VLF submarine communication transmitters that operate near ~20 kHz are calculated for comparison with the previous conflicting results [e.g., *Jean et al.*, 1960; *Steele and Chilton*, 1964]. The phase propagation velocities are inferred from the time delay of the radio signals received at different locations and the corresponding distance differences between different propagation paths from the transmitters to the receivers. The main benefit from studying VLF transmissions is their well-known operation [e.g., *Thomson*, 2010] and the locations of the radio transmitters [e.g, *Füllekrug et al.*, 2015, supplement].

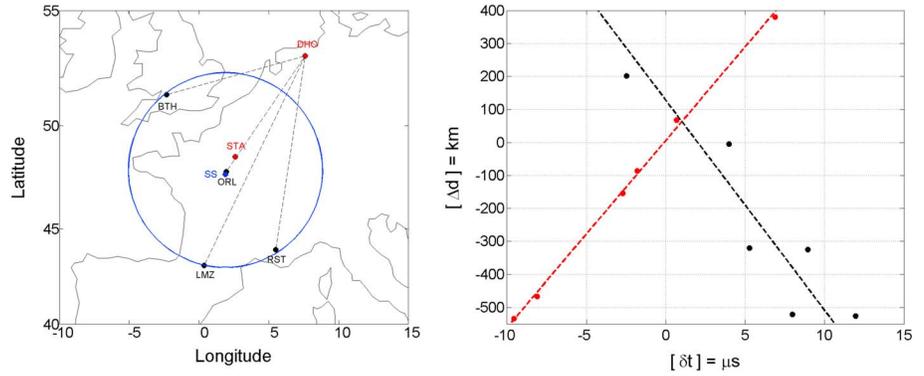

**Figure 1.** The geometry of the experiment and the measured phase velocities inferred from two VLF transmitters. (left) The locations of the two VLF transmitters (red dots) and the long-baseline (>500 km) radio receivers (black dots) used in this study. The circle (blue line) that determines the inside and outside of receiver array is centered on the sweet spot (blue dot) of receivers. (right) The stable ratio of distance differences and the measured time differences between the transmitters and the radio receiver pairs indicate a fixed phase propagation velocity that is larger than the speed of light for DHO (black) and smaller than the speed of light for STA (red). See text for details.

In this study, the transmissions from the French VLF transmitter in Sainte Assise (STA), operating at 20.9 kHz, and the German VLF transmitter (DHO) in Rhauderfehn, operating at 23.4 kHz, during a quiet period at night are used. These transmissions are chosen because their operating frequencies are close to the frequency range of the experimental lightning location system (5–15 kHz) and also because the transmitters are located inside and outside of the radio receiver array (Figure 1, left).

## 3.1. Instantaneous Phase Extracted From the Complex Trace

The time-dependent radio signal is treated as an analytic signal or complex trace. This allows the determination of the instantaneous phase at each sample in time series. The complex trace can be obtained from the real-valued recordings of the vertical electric field strength using the Hilbert transform [e.g., *Taner et al.*, 1979; *Schimmel and Paulssen*, 1997]. The Hilbert transform can be thought of as the convolution of the real signal with $1/(\pi t)$, where $t$ is time. This convolution generates the Hilbert transform as an output of a linear time-invariant system using the Cauchy principal value to avoid singularity. The Hilbert transform $H(f(t))$ of a real-valued time-dependent function $f(t)$ can also be understood as the effect of a phase shift of the negative frequencies by +90° and positive frequencies by −90°, where the multiplication with the imaginary unit $j$ to calculate the complex trace $F(t)$ shifts the negative frequencies by another +90° and restores the positive frequencies

$$F(t) = f(t) + jH(f(t)) = A(t)e^{j\varphi(t)},$$ (1)

where $A(t)$ is the time-dependent amplitude and $\varphi(t)$ is the time-dependent phase.

In practice, the complex signal can be calculated from the real signal by doubling the positive frequencies of $f(t)$ and by eliminating the negative frequencies of the real signal. This complex trace is subsequently down converted by multiplication with $e^{-j\Delta\omega t}$ to reference the phase to the start time of the recordings. The term $e^{-j\Delta\omega t}$ is the frequency shift operator that centers the spectrum at zero frequency, where $\Delta\omega$ is the shift frequency. The frequency shift operator has the in-phase and quadrature components, so this step can alternatively be used for obtaining a complex signal. A low-pass filter is applied to the down-converted signal to extract the target frequency band, which is normally a 150 Hz single-sided bandwidth for VLF transmitters. The phase of the processed signal, known as the instantaneous phase, can be extracted for each individual sample

$$\varphi(t) = \tan^{-1}\left[\text{Imag}\left(F(t) * e^{-j\Delta\omega t}\right) / \text{Real}\left(F(t) * e^{-j\Delta\omega t}\right)\right].$$ (2)

Cyclic ambiguities occur when the propagation path is longer than the wavelength of the radio wave. To avoid such ambiguities, the recorded waveforms are shifted by the propagation time from the transmitter to the receiver with an assumed propagation velocity. The shifted waveforms are calculated here by multiplication with the time shift operator $e^{-j\omega\Delta t}$ in the frequency domain, where $\Delta t$ is the propagation time. In first-order approximation, the propagation time is $\Delta t = d/c$, where $d$ is the propagation distance and $c$ is the assumed propagation velocity at the speed of light.

### 3.2. Phase Propagation Velocity Calculation

The phase propagation velocity is $v = d/t$, where $d$ is distance and $t$ is time. Assuming that the velocities are the same for different propagation paths in first-order approximation, the velocity can be determined as $v = \Delta d/\Delta t$, where $\Delta d$ is the distance difference between two propagation paths, i.e., $d_1 - d_2$, and $\Delta t$ is the time difference that results from the wave propagation, i.e., $\Delta d/c + \delta t$. The phase propagation velocity $v$ for one pair of receivers can then be inferred from

$$v = \frac{\Delta d}{\dfrac{\Delta d}{c} + \delta t} \, , \tag{3}$$

where $\delta t = \delta\varphi/\omega$ is the time residual calculated from the observed phase residual $\delta\varphi$ measured with respect to the speed of light, i.e., after shifting the two down-converted complex traces, and $\omega$ is the angular carrier frequency of the radio wave. The signs of these differences need to be consistent. The time residual $\delta t$ of the shifted waveform between pairs of receivers will be 0 if the radio waves propagate exactly at the speed of light, i.e., $v = c$.

The time residuals between pairs of receivers for the two transmitters depend on the distance differences between the transmitters and the receivers (Figure 1, right). Equation (3) can be reformulated to determine the ratio between the distance and time differences $\Delta d/\delta t = v \times c/(c - v)$. A fixed ratio between $\Delta d$ and $\delta t$ corresponds to a fixed velocity for the radio wave propagation. The slope of the linear relation between $\Delta d$ and $\delta t$ shows that the averaged phase propagation velocities are practically constant for each of the transmitters. The phase propagation velocities are ~0.64% faster and ~0.51% slower than the speed of light for the transmitters in Rhauderfehn (DHO) and Sainte Assise (STA), respectively.

This result highlights that it is possible for the phase propagation velocity of a radio wave to be smaller or larger than the speed of light. For example, a phase propagation velocity larger than the speed of light can occur when the radio wave arrives at an array of radio receivers with a certain elevation angle [e.g., *Füllekrug et al.*, 2015, Figure 2, right; *Rost and Thomas*, 2002, Figure 1]. In this case, the radio waves from Rhauderfehn would have a larger phase propagation velocity possibly because the radio waves arrive from larger elevation angles when compared to the transmitter Sainte Assise located inside the radio receiver network. The received VLF transmission consists of ground and sky waves as a result of waveguide propagation effects. This result strongly suggests that the ground wave is more attenuated for Rhauderfehn than Sainte Assise since it is located at a larger distance from the network (compare to Figure 1, left, and *Mezentsev and Füllekrug* [2013, Figure 8]). It follows that the phase propagation velocity depends on the location of the radiation source and the ground parameters along the propagation paths to the radio receivers. This result compares well with results previously reported in the scientific literature [e.g., *Jean et al.*, 1960; *Steele and Chilton*, 1964]. Thus, small lightning location uncertainties, caused by the mixture of ground wave and sky waves as a result of waveguide propagation effects, can be introduced if a fixed value for the wave propagation velocity is used in a TOA lightning location system.

## 4. Lightning Location Method and Simulation

The long-baseline lightning location system reported here uses a TOA approach in the frequency range 5–15 kHz. This frequency range is chosen because the return stroke deposits most of its energy there such that the phase is well determined. The signatures from lightning discharges are identified in the recordings, and the time differences between pairs of radio signals are determined to calculate the best possible lightning location. In theory, the lightning location is the intersection between the hyperbolas of constant distance differences that define possible lightning locations. The hyperbola for each pair of receivers can be determined from the distance differences between two receivers by multiplication of the measured time of arrival differences with a predetermined fixed wave propagation velocity [*Dowden et al.*, 2002].

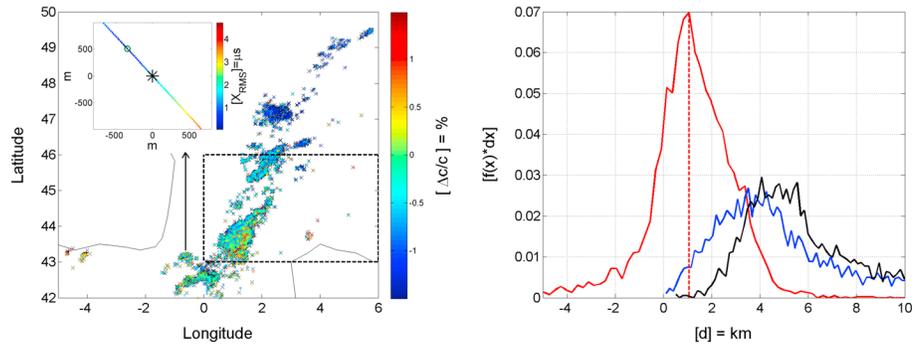

**Figure 2.** The lightning locations inferred from variable phase propagation velocities and comparison with ATDnet locations reported by the UK MetOffice. (left) The lightning locations inferred from a variable phase propagation velocity (colored circles) are compared with locations inferred from a fixed velocity (grey crosses) during 1 h of recording. The deviations from a propagation at the speed of light exhibit a smooth spatial change (color). The thunderstorm moved northeastward where numerous lightning discharges occurred and which are used to map the phase propagation velocity (dashed area). (inset) The calculated lightning location moves from northwest to southeast when the propagation velocity is varied from 0.995c to 1.005c, where the color indicates the RMS value of the time differences. The location calculated with the speed of light (cross) is 700 m away from the location with the minimum RMS (circle). (right) The lightning locations inferred from the variable velocity (blue line) and fixed velocity (black line) are compared with lightning locations reported by the UK MetOffice. The comparison indicates an improvement of the average location accuracy by ~890 m (red line) when using the variable phase propagation velocity.

The arrival time of the lightning return stroke signal is defined by the first peak, i.e., the nearest local maximum that precedes the absolute maximum in the complex trace in the filtered data. Thus, the time differences between stations are taken as the difference between the corresponding arrival times of the return stroke. This method is used as it is found to be the best compromise between location accuracy and computational effort. Using the first peak as the arrival time should, in theory, exclude the effect of the sky waves in the transient lightning signal. In practice, it is found that there is still some effect due to the digital filtering. For example, the average distance from the lightning to the receivers is about ~560 km in the recorded 9 h. The time delay between the ground wave and sky wave is about 95 μs making a simplified assumption that the ground is flat and the ionosphere is at 90 km in height. The rise time of the return stroke is ~5 μs, and decay time to the half-peak value is 70 to 80 μs [*Rakov*, 2013]. The narrowband filter broadens the two separate impulses in time such that they overlap slightly and the peak from the ground wave contains some contribution from the sky wave.

In practice, the TOA method is to find the closest match between the measured time differences and the time differences associated with the best possible lightning location. This closest match is determined here by minimizing the root mean square (RMS) value

$$X_{\text{RMS}} = \sqrt{\frac{1}{N}\sum_{n=1}^{N}\left(\Delta t_n - \frac{\Delta d_n}{v}\right)^2}, \tag{4}$$

where $N$ is the number of receiver pairs, $\Delta t_n$ are the measured time differences between receivers, $\Delta d_n$ are the distance differences from the best possible lightning location to the receivers, and $v$ is the wave propagation velocity which defaults to the speed of light in many lightning detection systems.

The wave propagation velocity used in equation (4) influences the determination of the lightning locations. As a result, different lightning locations and their corresponding RMS values for several lightning events are calculated for different velocities. For example, one typical lightning discharge occurred inside the lightning location system in southern France (latitude 43.6929°, longitude 0.6077°) at 18:01:31 and 189,486 μs on 8 August 2014. The propagation velocity is varied in steps of 0.01% within ±0.5% of the speed of light. The variation of the lightning location and RMS value demonstrates the importance of the propagation velocity in the lightning location system (Figure 2, left, inset).

The lightning location shifts gradually from northwest to southeast with increasing propagation velocity. The velocity with the minimum RMS value is 0.17% slower than the speed of light. The distance between the location with the minimum RMS value and the location calculated with the speed of light is ~700 m. This suggests that a fixed wave propagation velocity can introduce hundreds to thousands of meters difference in the lightning location determination if the radio waves from lightning discharges do not propagate at the speed of light.

## 5. Lightning Locations Inferred From Variable Phase Propagation Velocity

The phase propagation velocities of the VLF radio waves are slightly different from the speed of light as demonstrated by the observed phase propagation velocities of the VLF transmissions for submarine communication. In order to reduce the location uncertainty caused by the use of a fixed velocity, a variable phase propagation velocity is implemented in the lightning location system. This step is based on the idea that the phase propagation velocities for different wave propagation paths vary as a result of sky wave contributions and ground effects. The calculated locations with variable phase propagation velocities are possibly closer to the real lightning location because the real wave propagation is more complex than light traveling through a vacuum. The lightning discharges inferred from variable phase propagation velocities and fixed velocities are now compared for many events.

At least three independent time differences are necessary to enable the unique determination of three independent parameters, i.e., latitude, longitude, and the phase propagation velocity. This calculated phase propagation velocity represents an averaged phase propagation velocity for the 5–15 kHz frequency range along the propagation paths from that particular lightning location to all the radio receivers. The phase propagation velocity calculated from the optimization of the RMS function in equation (4) needs to be constrained because we know that the propagation velocity cannot be largely different from the speed of light even if it differs slightly from the speed of light. The phase propagation velocities inferred from radio waves of lightning discharges are calculated using the data recorded from 18:00 to 19:00 (UTC) on 8 August 2014. About ~68% ($\pm1$ $\sigma$) of the calculated phase propagation velocities do not exceed $\pm1.5\%$ of the speed of light. Thus, the results with a phase propagation velocity outside this range are considered to be questionable, possibly as a result of interference, such that these phase propagation velocities are not used for further analysis. More than 80% of the calculated phase propagation velocities are slower than the speed of light, possibly because most of the contribution from sky waves is mitigated by the time difference selection method described in section 4. The lightning locations inferred from fixed velocities and variable phase propagation velocities in France are roughly similar on a large scale (Figure 2, left).

In order to assess the performance of this novel lightning location method, the locations calculated by the variable phase propagation velocity and the fixed velocity are compared with the lightning locations reported by the commercial lightning location system ATDnet of the UK MetOffice (Figure 2, right). ATDnet is the first lightning detection network used in Europe [*Lee*, 1986; *Lewis et al.*, 1960]. It has more radio receivers, and the receivers are more evenly distributed across Europe than the experimental lightning location system investigated here, such that the results of ATDnet can be taken as a ground truth. The distances between the lightning locations inferred from variable phase propagation velocities and the lightning locations reported by ATDnet are smaller than the distances between the locations inferred from the speed of light and the locations reported by ATDnet. The average improvement of the distances to ATDnet locations is ~890 m, and the most likely improvement of the distances is ~1.06 km. A similar result is inferred from a comparison with the locations reported by the lightning location system Meteorage. By using variable phase propagation velocities, the calculated results are ~0.78 km closer to Meteorage locations. Meteorage uses shorter distances between receivers such that the contributions of sky waves to the observed radio waves are negligible compared to long-range lightning detection networks. The improvement of the lightning locations by ~0.89–1.06 km therefore indicates that the use of variable phase propagation velocities for lightning location mitigates the effect of sky wave contributions and ground effects.

## 6. Velocity Map

The calculated phase propagation velocities in the given frequency range are very similar in neighboring areas and exhibit a distinct smooth change over larger areas (Figure 2, left). This observation leads to the idea of creating a velocity map to characterize the mitigation of sky wave contributions and ground effects in a larger area. The velocity map uses one representative phase propagation velocity in a small area if the observed phase propagation velocities in this area fit a normal distribution well.

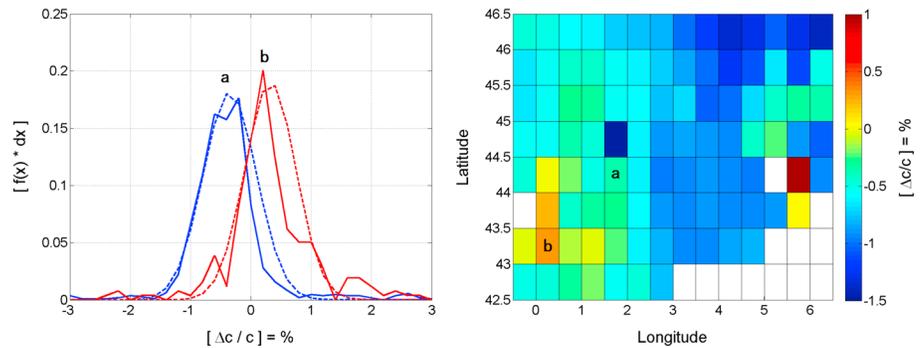

**Figure 3.** The velocity map and the distributions in two grid cells of the map. (left) The velocity distributions (solid lines) of two sample grid cells compare well to normal distributions with small standard deviations (dashed lines). One grid cell extends from 44° to 44.5°N and 1.5° to 2°E (red line), and the second grid cell extends from 43° to 43.5°N and 0° to 0.5°E (blue line). (right) The velocity map is composed of individual grid cells that exhibit a smooth spatial gradient. The white grid cells have fewer than 40 lightning discharges and are therefore not shown.

Radio signals from more than 30,000 lightning discharges are recorded within ~9 h in an area extending from 0° to 6°E and from 43° to 46°N on 8 August 2014. The area of interest is divided into grid cells of 0.5° in latitude and longitude for analysis. Note that the grid cells are not square in distance because the latitudes are not great circles. This grid cell size was chosen as a trade-off between spatial resolution and getting large enough sample populations of lightning discharges in individual cells.

The statistical stability of the phase propagation velocities in a grid cell is tested before the calculation of the velocity map. The velocity distributions in two example grid cells fit different normal distributions well (Figure 3, left). The mean velocities for these two grid cells are $0.9965c$ and $1.0033c$ with standard deviations of $0.0044c$ and $0.0042c$, respectively. The mean velocities therefore differ significantly from the speed of light. The distributions within the example grid cells demonstrate that the mean phase propagation velocity appropriately represents the distributions of phase propagation velocities in a small region.

The mean phase propagation velocities for each individual grid cell are different (Figure 3, right). A few grid cells (shown in white) contain fewer than 40 events—too few to support a statistically meaningful result. For two grid cells with poor statistical distributions, i.e., grid cells from 44° to 44.5°, 5.5° to 6°E, and 44.5° to 45°N, 1.5° to 2° E, the median phase propagation velocities are unusually small and large, respectively, and are hence excluded from further analysis. For most grid cells, the phase propagation velocities are smaller than the speed of light by ~0.1–1%. There are also some locations that show a mean phase propagation velocity larger than the speed of light. This observation indicates that these radio waves do not arrive from the horizon at zero elevation angles, similar to the radio transmissions described in section 3.

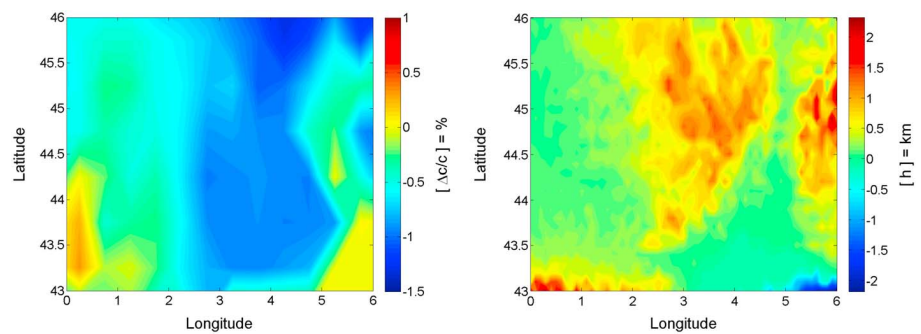

**Figure 4.** Comparison of the final velocity map with a map of the topographic elevation. (left) The final velocity map is calculated from the interpolated contours of the individual grid cells, and the map exhibits a smooth spatial change. (right) The topographic elevation map shows the Massif Central in the center of France and exhibits a similar pattern when compared to the final velocity map. The area with high elevation represents mountains, and the area with the negative elevation is the seabed of the Mediterranean south of France.

Assuming that the mean phase propagation velocity for each grid cell represents the phase propagation velocity in the given frequency range for the center location, contour lines can be drawn between the centers to produce a smoothed velocity map (Figure 4, left). The phase propagation velocities in the two invalid grid cells are replaced by the mean values of their surrounding grid cells. The velocities of the grid cells with too few samples are filled with the median value of all the other phase propagation velocities. The final map shows a smooth variation of the phase propagation velocities across the studied area.

## 7. Discussion and Conclusion

The comparison of the inferred lightning locations with the lightning locations reported by ATDnet and Meteorage as ground truth strongly suggests that the lightning locations inferred from the variable phase propagation velocity are more accurate than the locations inferred from the fixed velocity in this VLF long-range lightning location system. This increase of accuracy is attributed to the ability of the variable phase propagation velocity to mitigate the influence of sky waves and ground effects on the calculation of the lightning location. It is interesting to note that ATDnet uses a fixed velocity for the calculation of their lightning locations. It therefore appears plausible that these lightning locations could also be improved by use of variable velocities to mitigate sky wave contributions but perhaps to a smaller degree because ATDnet uses more radio receivers. For yet more resource intensive short-range lightning location networks that use mainly ground waves to determine lightning locations, variable velocities could only mitigate wave propagation effects associated with variations of the ground conductivity and terrain.

The velocity map shows a smooth spatial variation of the phase propagation velocities from radio waves of lightning discharges in central and southern France. This velocity map represents the propagation velocities of radio waves from lightning discharges as they appear to the lightning location system. The map of phase propagation velocities therefore reflects the average impact of sky waves and ground waves on the calculation of lightning locations as a result of the network configuration, i.e., the different propagation paths between the lightning locations and the radio receivers. It is conceivable that the map of phase propagation velocities is a combined product of ground wave propagation influenced by the ground conductivity and sky wave propagation influenced by ionospheric conductivity. In this case, a phase propagation velocity map could be derived with a denser network of radio receivers and longer recordings to reflect other geophysical properties such as the topography and ground conductivities [*International Telecommunication Union*, 2015]. To elucidate this speculative possibility, the map of phase propagation velocities is compared to a terrain elevation map in France (Figure 4, right). Interestingly, it can be observed that the phase propagation velocities of lightning discharges at higher altitudes are smaller than at lower altitudes. For example, the phase propagation velocities over the mountainous area of the Massif Central are slightly slower than the speed of light, possibly because the mountain peaks disturb radio wave propagation. This result seems to confirm recent modeling work that suggests that the time delay of lightning radiated electromagnetic fields can be significantly affected by the presence of mountainous terrain [*Li et al.*, 2016a]. The phase propagation velocity in the flatlands can occasionally exceed the speed of light, which indicates the arrival of the radio wave from an elevation angle. This result might possibly be due to the elevation profile or the dispersion caused by the ground conductivity profile along the propagation path and is subject to further research. Note that the phase propagation velocity is inferred from differential measurements between receiver pairs; therefore, the expectation is that the geophysical property around source might not be a dominant factor, which possibly explains why the correlation is poor in some grid cells. Nevertheless, it is also possible that the properties of the propagation path near the source have a larger effect on the total phase delay than the properties far away from the source. This is suggested, for example, by the Sommerfeld-Norton attenuation function over a flat Earth [e.g., *Galejs*, 1972, Figure 9.3] with a similar effect inferred for an inhomogeneous conductivity of the Earth [e.g., *King et al.*, 1966]. It therefore appears to be promising to compare various phase propagation velocity maps inferred from other lightning detection networks to identify the influence of geographic features on VLF radio wave propagation [*Barr et al.*, 2000].

The relationship between the phase propagation velocity and the ionospheric conditions is another promising area for future research. The observed phase propagation velocities of the transmissions were different when thunderstorms occurred along the propagation path. These ionospheric conditions also change with time, e.g., the diurnal variability discussed in *Schonland et al.* [1940]. The studied thunderstorm propagated eastward during 9 h from day to night. As a result, characteristics of the wave propagation in different ionospheric conditions may be revealed by analyzing the calculated phase propagation velocities at different times of day. This is planned for future work.

In summary, this study enabled several results. (1) The phase propagation velocities inferred from radio waves emitted by VLF transmitters can be smaller or larger than the speed of light. (2) Simulations show that lightning locations calculated with different phase propagation velocities can cause deviations of the lightning location by hundreds to thousands of meters when compared to the location inferred from the speed of light. (3) A long-range lightning location system that uses variable phase propagation velocities can improve the lightning location accuracy by ~0.89–1.06 km when compared to lightning locations inferred from an array with more radio receivers. (4) As a result of the network configuration, the phase propagation velocities are mapped over central and southern France to summarize the impact of sky waves and ground effects on the calculation of lightning locations.

**Acknowledgments**
The work of Z.L. is sponsored by
the University of Bath, UK MetOffice
(EA-EE1077) and the China Scholarship
Council (CSC) (File 201408060073). The
work of K.K. is sponsored by the
Engineering and Physical Sciences
Research Council (EPSRC) under DTA
contract EB-EE1151. The work of M.F.
and A.M. is sponsored by the Natural
Environment Research Council (NERC)
under grants NE/L012669/1 and
NE/H024921/1. The data used for this
publication is available from http://doi.
org/10.15125/BATH-00242. The eleva-
tion data used in Figure 4 (right) was
downloaded from NOAA, available
online at https://www.ngdc.noaa.gov/
mgg/global/global.html, and
Meteorage data were kindly provided
by Torsten Neubert through the
European Science Foundation network
TEA-IS. Z.L. wrote the paper and per-
formed the data analysis, K.K. advised
on the use of the VLF transmitters and
the Hilbert transform, A.M. assisted with
the installation of the receiver network,
S.E. and J.S. helped with the interpreta-
tion of the results, and M.F. supervised
the work of Z.L. and advised on the
concepts for the data analyses. The
authors wish to thank Serge Soula, Jean-
Louis Pincon, Stephane Gaffet, and their
teams for hosting the radio receivers in
Lannemezan, Orleans, and Rustrel. Z.L.
wants to thank Dirk Klugmann and Ivan
Astin for inspiration and encourage-
ment toward this project and acknowl-
edges helpful discussions and support
from Andrew Moss, Nathan Smith, Neil
Hindley, and Corwin Wright, as well as
the constructive suggestions of the
anonymous reviewers.